\documentclass[aps,twocolumn,prl]{revtex4}
\usepackage{graphicx}
\usepackage{latexsym}
\usepackage{amsmath}
\usepackage{amsfonts}
\usepackage{amssymb}

\begin{document}

\title{Thermodynamic restrictions on evolutionary optimization 
of transcription factor proteins}

\author{Longhua Hu and Alexander Y. Grosberg}
\affiliation{Department of Physics, University of Minnesota, 116
Church Street SE, Minneapolis, Minnesota 55455, USA}

\author{Robijn Bruinsma}
\affiliation{Department of Physics and Astronomy, University of California at Los Angeles, 
Los Angeles, California 90035}

\begin{abstract}

\textit{Conformational fluctuations} are believed to play an important role
in the process by which transcription factor proteins locate and bind their
target site on the genome of a bacterium. Using a simple model, we show that
the binding time can be minimized, under selective pressure, by adjusting 
the spectrum of conformational states so that the fraction of time spent in
more mobile conformations is matched with the target recognition rate. 
The associated optimal binding time is then within an order of magnitude
of the limiting binding time imposed by thermodynamics, corresponding to
an idealized protein with instant target recognition. Numerical estimates
suggest that typical bacteria operate in this regime of optimized 
conformational fluctuations.

\end{abstract}

\maketitle

The ability of bacteria to respond within minutes to changes in their environment relies
on genetic switches that are controlled by \textit{transcription factors}. Transcription
factors are proteins that---following activation by an environmental change---are able to locate
a specific region (the ``operator sequence") along the bacterial genome and bind to it,
thereby regulating the expression of a gene (or group of genes) adjacent to that region 
\cite{Alberts}. The number of copies of a transcription factor protein associated with a
specific gene varies, but typically it is in the range of $10^2$. Because bacterial genomes
have a size in the range of $10^7$ sites, a transcription factor must be able to ``scan"
the DNA for the target site at a rate of $10^5$ sites per second or faster in order for at
least one of them to reach the target site within seconds. Note that following the \textit{search}
for the target site, the transcription factor still has to \textit{bind} to the target site 
to regulate the expression of the gene.

A series of classical papers on the search process \cite{Adam_Delbruck, Riggs, Richter}
culminated in the work of Berg, Winter and von Hippel (BWH) who showed \cite{BWH}---for
the canonical case of the \textit{lac} repressor protein of the bacterium \textit{E. coli}
---that the search process takes place not by straightforward 3D diffusion to the target
binding site but rather by a slide-jump combination of 1D diffusional sliding along the
DNA chain alternating with 3D diffusional jumps between different DNA segments. By
restricting part of the search to the 1D ``target space", the binding rate is effectively
enhanced with respect to a pure 3D search, while the 3D jumps reduce the repetitive visits
to the same sites that characterize purely 1D diffusive searches. This scenario is made 
possible by a modest, non-specific electrostatic affinity between the transcription factor
and duplex DNA. BWH also provided evidence that, under physiological conditions, the search
time has a minimum with respect to the strength of this non-specific affinity, which
may be the result of evolutionary optimization under selective pressure. Subsequent structural
studies \cite{Kalodimos} have shown that the DNA-binding domains of the \textit{lac} repressor
are subject to strong \textit{conformational fluctuations} when the protein is in contact
with non-operator DNA. If the binding domain is in contact with operator sequence DNA
then the protein can undergo a large-scale conformational change to a stable structure with
direct contacts between the amino-acid side chains and the DNA bases.

It would seem obvious that the delay time between activation and binding of a transcription
factor to the operator sequence (``binding time") is minimized by maximizing the 1D diffusion
constant $D_1$. However, simply increasing the transport rate will impair the accuracy, or 
\textit{fidelity}, with which the protein can distinguish a right from a wrong site. Specifically,
if the binding of a transcription factor to the target site is characterized by a certain rate
$\Omega$, then the protein is likely to \textit{overshoot} the target site if the jump rate
$D_1/a^2$ between sites, with $a$ the spacing between protein binding sites, is large compared
to $\Omega$. Similar conflicts between process speed and process fidelity are familiar from
DNA duplication and transcription where increased reaction rates increase the number of 
duplication and transcription errors. 

\begin{figure}
\centerline{\scalebox{0.25}{\includegraphics{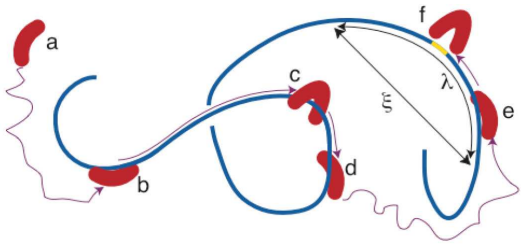}}}
\caption{(Color online) Schematic representation of the model. A protein moving diffusively
through the cell volume (a) is adsorbed on genomic DNA (b) where it adopts one of two
conformations: $+$ and $-$. In the $+$ conformation it is loosely associated with
the DNA and can move by one-dimensional diffusion along the DNA chain (b) while in the $-$ 
conformation (c) it is tightly associated with the DNA and is immobile. After
returning to the $+$ state, it restarts the sliding motion. The protein also can desorb
from the chain (d) and return to three-dimensional diffusive motion. Following a number
of such cycles, the protein lands in the ``antenna region" within a distance $\lambda$ 
of the target binding site (e). After reaching the target site by one-dimensional diffusion
it can undergo a large-scale irreversible conformational transition to the final
bound state if it is in the $-$ state (f).}
\label{fig:Cartoon}
\end{figure}

Slutsky and Mirny \cite{Mirny} proposed that conformational fluctuations could ease
the conflict between speed and fidelity. If some conformations of the transcription
factor are sensitive to the DNA sequence while others are characterized by rapid 
transport then the transcription factor might be able to scan the genome efficiently
by rapidly flipping between the two types of conformations.
The aim of this paper is to analyze how close this mechanism can approach limits
of search efficiency imposed by fundamental principles of thermodynamics.
We will address this question by examining a simple model for the conformational
fluctuations, similar to that of Ref. \cite{Mirny}, where the transcription factor
is allowed to adopt only two conformations ($+$ and $-$) when in contact with
non-operator DNA. As illustrated in Fig. \ref{fig:Cartoon}, in the $+$ state,
the protein is less ordered and only loosely associated with the DNA while
it can slide along the DNA chain. In the $-$ state, the protein is more ordered,
closely associated with the DNA and immobile \cite{Wolynes}. If the transcription
factor is in contact with the target operator sequence then, in addition to these
two states, it also can undergo an irreversible conformational transition from
the $-$ state to the fully ordered final bound state. We will show that the 
\textit{shortest possible binding time} in this model is controlled by a dimensionless
binding rate $\omega \equiv 2\Omega ab /\sqrt{KD_1 D_3}$, with $D_3$ the protein diffusion
coefficient in bulk solution, $D_1$ the diffusion coefficient for 1D transport 
along the DNA in the $+$ state, $K$ the equilibrium constant for the non-specific 
protein-DNA interaction, and $b$ the DNA-protein ``capture radius" \cite{Radius}. 
If the dimensionless binding rate is comparable to one---or larger than one---, 
then we can show that for a particular value of the energy difference $\Delta E_{\pm}$
between the $+$ and $-$ conformations, the binding time can approach an absolute lower
bound that corresponds to proteins having \textit{infinitely fast} final binding rates.
In other words, if the internal degrees of freedom of the protein in the sliding state
are properly matched to the final binding rate then the binding time of a transcription
factor can approach the shortest possible value allowed by thermodynamics \textit{provided}
the dimensionless binding rate is sufficiently large.

To demonstrate these claims, assume a cell of volume $V$ containing a DNA genome of
length $L$. The cell also contains a certain (low) concentration $c$ of transcription
factor proteins that can bind reversibly and non-specifically to the DNA. A protein whose
center is located inside a cylindrical tube of radius $b$ surrounding the duplex DNA will
be assumed to be non-specifically associated with the DNA. The fraction $\phi$ of the
total cell volume occupied by the tube is of the order of $Lb^2/V$. There is also a single
target site on the strand where the transcription factor can bind irreversibly. We start
by applying a fundamental theorem \cite{Pontryagin}, which---in terms of our model--- 
states that the mean waiting time for irreversible occupation of the target site, the 
quantity of interest to us, is equal to the inverse of a steady-state diffusion current of
a \textit{different} problem, namely one where the target site is replaced by a protein 
\textit{sink} that constantly absorbs $-$ state transcription factors located at the target 
site at a rate $\Omega$, while the protein concentration far from the target site is
maintained at a certain fixed value $c_3({\infty})$. The steady-state diffusion current, 
denoted by $J_{\rm 3D}$, into the target site for this second problem can be obtained from
straightforward solution of the diffusion equation, which leads to the well-known Smoluchowski
relation for the reaction rate of diffusion-limited chemical reactions:
\begin{equation}
J_{\rm 3D} \sim D_3 c_3(\infty) \xi \ .
\label{eq:Smoluchowski}
\end{equation}

Following Ref. \cite{Grosberg}, the effective ``target radius" is defined as the radius
of a sphere, surrounding the target site, that determines a cross-over regime such that
far outside the sphere adsorption of proteins onto the DNA chain is in equilibrium with
evaporation of protein from the DNA chain while deep inside the sphere the absorption rate
exceeds the evaporation rate. For the case of transcription factors obeying BWH slide-skip
transport, the size of this target sphere is determined by the condition that if a protein
lands on a DNA segment inside the target sphere, following a 3D diffusion step, then
it typically reaches the target sink by pure 1D diffusion where it gets absorbed before
there is a chance for it to ``evaporate" and leave the DNA. The length $\lambda$ of DNA
chain inside this target sphere---referred to as the ``antenna" length---in general depends
on the spatial organization of the genome. We will assume here the simple case of a straight
genome, with $\xi$ of order $\lambda$ \cite{LambdaP}. This antenna length has to be determined
self-consistently but first we must establish a relation between $c_3(\infty)$ and 
the actual protein concentration $c$.

Far outside the target sphere the DNA-protein system is, by assumption, nearly in local
thermal equilibrium, so one can determine the concentrations of adsorbed and free
proteins purely from equilibrium considerations. If one views the association of
the transcription factors with DNA as a simple chemical reaction, then
the concentration $\tilde c(\infty)$ of proteins adsorbed non-specifically on the DNA
and the concentration $c_3(\infty)$ of free proteins must be related to the reaction
volume fraction $\phi$ by the \textit{Law of Mass Action} for dilute chemical systems
in thermodynamic equilibrium:
\begin{equation}
\frac{c_3(\infty) \phi} {\tilde c(\infty)} \simeq K 
\label{eq:MassAction}
\end{equation}
with $\phi \ll 1$. The non-specific protein-DNA equilibrium constant $K$ depends
strongly on the salt concentration \cite{Richter}, and other thermodynamic parameters,
but it is independent of the protein and DNA concentrations. Since 
$c = c_3(\infty) + \tilde c(\infty)$, the concentrations of free and adsorbed proteins
are now determined but it will be useful to replace the bulk concentration $\tilde c(\infty)$
of adsorbed proteins by the 1D concentration $c_1(\infty) \simeq b^2 \tilde c(\infty)/\phi$, 
the number of adsorbed proteins \textit{per unit length} of DNA far from the target site.
Solving for $c_1(\infty)$ and $c_3(\infty)$ gives $c_1(\infty) \simeq cb^2/K(1+\phi/K)$
and $c_3(\infty) \simeq c/(1 +\phi/K)$, still for $\phi \ll 1$.

Deep inside the target sphere, the system is not in thermal equilibrium, with the adsorption
rate of proteins from the bulk solution to the DNA exceeding the evaporation rate. The 
difference is matched by a 1D diffusion current $J_{\rm 1D}$ along the DNA chain towards
the target site. In order to estimate this 1D diffusional transport, note that if the
interconversion rate between the $+$ and $-$ states is sufficiently rapid then their
respective occupancies can be approximated by the equilibrium Boltzmann distribution. 
The \textit{effective} 1D diffusion constant for transport along the chain---which 
we will denote by $\tilde D_1$---is then proportional to the Boltzmann probability 
$p(+)$ to find the protein in the $+$ state. If $\mu \equiv \exp\left(-\Delta E_{\pm}/k_B T\right)$,
then $p(+) = \mu/(1 + \mu)$ and $\tilde D_1 \simeq D_1 \mu/(1 + \mu)$. Similarly, the 
\textit{effective} target site binding rate $\tilde \Omega$ is, under these same conditions,
proportional to the probability $p(-) = 1 - p(+)$ to find the protein in the $-$ state
and $\tilde \Omega \simeq \Omega/(1 + \mu)$. 

Let $c_1(0)$ be the 1D concentration at the target site. If the final binding rate
were infinitely fast, then $c_1(0)$ would be zero but, because of the overshoot effect,
this is no longer the case. If we view the surface of the target sphere as a matching
region between the asymptotic regions far from the sink where the 1D concentration 
approaches $c_1(\infty)$ and the region deep inside the target sphere near the sink 
where the 1D concentration approaches $c_1(0)$, then we can estimate the 1D concentration
gradient as $\left[c_1(\infty) - c_1(0)\right]/\lambda$. It follows that the 1D diffusion
current towards the sink equals:
\begin{equation}
J_{\rm 1D} \sim \tilde D_1 \frac{ c_1(\infty) - c_1(0) } {\lambda} \ .
\end{equation}

The number of proteins absorbed per second by the sink itself, $J_{\rm s}$, 
is of the order of $a c_1(0) \tilde \Omega$, with $a$ the spacing between
protein binding sites. Conservation of the number of proteins requires the
three currents $J_{\rm 3D}$, $J_{\rm 1D}$ and $J_{\rm s}$ to be equal to 
each other \cite{Grosberg}, so 
\begin{equation}
J_{\rm 3D} = J_{\rm 1D} = J_{\rm s} \ .
\label{eq:Conservation}
\end{equation}

Equating the 1D diffusion current with the sink current allows us to eliminate
$c_1(0)$ with the result:
\begin{equation}
J_{\rm 1D} \sim \frac{\tilde D_1 c_1(\infty)} {\lambda} 
\left( \frac{\tilde \Omega} {\tilde \Omega + \tilde D_1/a\lambda} \right) \ .
\end{equation}
The factor in front of the square brackets is the diffusion current in the absence
of overshoot. The importance of overshoot is thus determined by the dimensionless
number $a\lambda \tilde \Omega/\tilde D_1$. Since $\lambda^2/\tilde D_1$ is the 
typical time spent by a protein diffusing along the antenna, it follows that 
$a\lambda/\tilde D_1$ is the typical time spent near the target site so 
$a\lambda\tilde \Omega/\tilde D_1$ is the product of the typical time spent near
the target site with the effective absorption rate. The term inside the square
brackets can then be understood as the probability for a protein in the antenna 
region to be trapped by the target.

Equating the 1D and 3D currents provides us with a self-consistency condition that
determines both the size of the antenna length $\lambda$ and the reaction rate. 
Solving for $\lambda$ using Eqs. (\ref{eq:Smoluchowski}) and (\ref{eq:Conservation})
and using $c_1(\infty) / c_3(\infty) \sim b^2 / K$ gives the antenna length:
\begin{equation}
\lambda = \sqrt{ \frac{b^2} {K} \left( \frac{\tilde D_1} {D_3} \right) +
\left( \frac{\tilde D_1} {2 \tilde \Omega a} \right)^2 } 
- \frac{\tilde D_1} {2 \tilde \Omega a} \ .
\end{equation}
The maximum value, $\lambda_\infty = \sqrt{b^2 D_1/KD_3}$, is reached for infinite
$\Omega$ and infinite $\mu$.

\begin{figure}
\centerline{\scalebox{0.5}{\includegraphics{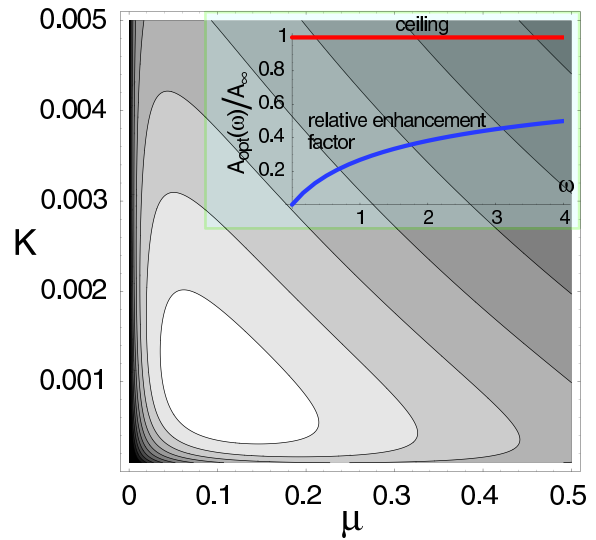}}}
\caption{(Color online) Contour plots of the transport enhancement factor $A$
as a function of the equilibrium constant $K$ and the occupation probability ratio
$\mu$ of the $+$ over $-$ states for $D_1 = 10^{-9}$ ${\rm cm}^2/{\rm s}$, 
$D_3 = 3 \times 10^{-7}$ ${\rm cm}^2/{\rm s}$, $a = 0.34$ nm, $b = 5$ nm,
$\phi=0.01$ and $\Omega= 3 \times 10^3$ Hz. There is a shallow maximum around
$\mu=0.1$ and $K=10^3$. The ratio of the transport enhancement factor at this maximum,
$A_{\rm opt}$, and the thermodynamic limiting enhancement factor $A_\infty$ equals
$0.193$. Inset: Dependence of the ratio $A_{\rm opt}/A_\infty$ on the dimensionless
binding rate $\omega$.}
\label{fig:Contour}
\end{figure}

It will be helpful to express the binding rate $J_{\rm 3D} \sim D_3 c_3(\infty) \xi$
in dimensionless units as $A \equiv J_{\rm 3D} /(c D_3 a)$ with $c D_3 a$ the 
Smoluchowski limiting rate of a conventional 3D diffusive search for an absorber
target of radius $a$ (the spacing between binding sites), so $A$ can be viewed
as a reaction \textit{amplification} or enhancement factor. This enhancement factor
can be expressed as a simple function of the dimensionless binding rate 
$\omega = 2 \Omega ab/\sqrt{K D_1 D_3}$ and the Boltzmann factor
$\mu = \exp\left(-\Delta E_{\pm}/k_B T\right)$:
\begin{equation}
A(\omega, \mu) \sim A_\infty \frac{1} {\omega} 
\left(\sqrt{\omega^2 \frac{\mu} {1+\mu} + \mu^2} - \mu \right) \ .
\end{equation}
Here $A_\infty = (b/a) \sqrt{KD_1/D_3} /(K + \phi)$ is the maximum value
of the enhancement factor, corresponding to $\lambda = \lambda_\infty$
with both $\mu$ and $\Omega$ infinite. We will examine the amplification factor
$A(\omega, \mu)$ as a function of the non-specific equilibrium constant $K$ and
the occupation ratio $\mu$ of the $+$ state and $-$ state, rather than $\omega$
and $\mu$, because these are physical parameters characterizing the interaction
between the transcription factor and the DNA that are expected to be sensitive
to specific point mutations of the transcription factor amino-acid sequence 
through their exponential dependence on binding and activation energies. The contour 
lines of constant $A$ as a function of $K$ and $\mu$ in Fig. \ref{fig:Contour}
show that there is a single, rather shallow maximum. The physical origin behind 
the maximum of $A$ with respect to $K$ is, as discussed earlier, the fact that
a combination of 1D and 3D diffusion minimizes the search time. By contrast,
the maximum of $A$ as a function of $\mu$ at $\mu_{\rm opt} = (\sqrt{1+2\omega} - 1)/2$
is surprising because it might have been expected that for sufficiently long DNA,
location of the target site \textit{always} should be the ``rate-limiting" step,
in which case the optimal choice for $\mu$ would be infinite since that maximizes
the effective 1D diffusion constant $\tilde D_1 = D_1 \mu/(1 + \mu)$. It can be
shown that the maximum with respect to $\mu$ actually is a form of \textit{impedance
matching} with the effective ``resistance" of the 1D diffusional search matched
with the effective resistance of the binding process.

If $\mu$ adopts the optimal value $\mu_{\rm opt} = (\sqrt{1+2\omega} - 1)/2$, 
then the ratio $A_{\rm opt}/A_\infty$ of the optimal rate amplification 
factor and its maximum value is a function only of the dimensionless rate
$\omega$:
\begin{equation}
\frac{A_{\rm opt} (\omega)} {A_{\infty}} = 1 - \frac{1} {\omega} 
\left( \sqrt{1 + 2\omega} - 1 \right) \ .
\end{equation}
The dependence of $A_{\rm opt}/A_\infty$ on $\omega$ is shown in the inset of 
Fig. \ref{fig:Contour}: $A_{\rm opt}$ is of the same order of magnitude as the
theoretical limit $A_\infty$ already for modest values of $\omega$. This 
demonstrates our central claim: it is \textit{possible} for the overall binding
rate of a transcription factor to approach the theoretical limiting value but 
\textit{only} by a suitable choice of $\mu$, and \textit{only} if the dimensionless
binding rate $\omega$ is of the order of one, or larger than one.

Are these two conditions realistic for typical transcription factors? Typical
values for the diffusion constants of bacterial transcription factors are
\cite{Xie, Austin} $D_1 \sim 10^{-9}$ ${\rm cm}^2/{\rm s}$ and 
$D_3 \sim 3 \times 10^{-7}$ ${\rm cm}^2/{\rm s}$. We can estimate the protein-DNA
reaction volume fraction $\phi$ for \textit{E. coli} by assuming it to be comparable
to the DNA volume fraction (about $1\%$). The equilibrium constant can then be 
determined from the relation $c_3(\infty) \simeq c/(1 +\phi/K)$ and the fact that 
it is known that about $10\%$ of the \textit{lac} repressor proteins of \textit{E. coli}
are in solution \cite{Huang}, which means that $K$ must be of the order of $10^{-3}$. 
If we assume $a$ to be equal to the base-pair spacing $0.34$ nm, and estimate 
$b$ as $5$ nm, then the dimensionless binding rate $\omega$ is of the order 
of $10^{-4} \Omega$ with the binding rate $\Omega$ expressed in Hz. A large-scale 
protein conformational change typically involves millisecond to microsecond time scales,
from which it follows that $\omega$ must lie in the range of $0.1$ to $100$. Note, 
from Fig. \ref{fig:Contour} that the optimal value for $K$ is close to $10^{-3}$ for
$\Omega$ in the kHz range. We conclude that the second condition can be satisfied under
typical conditions. Next, the optimal occupation ratio 
$\mu_{\rm opt} = (\sqrt{1+2\omega} - 1)/2$ is in the rage of $0.1$ to $10$ for $\omega$
in the range of $0.1$ to $100$. The corresponding optimal energy difference $\Delta E_{\pm}$
between the $+$ and $-$ states is then in the range of a few $k_B T$, with $\Delta E_{\pm}$
positive for $\omega < 4$ but negative for $\omega>4$. In either case, the structure of 
``optimized" transcription factors bound to non-operator DNA should be subject to strong
thermal fluctuations. As we saw, this is indeed the case of the \textit{lac} repressor
\cite{Kalodimos}, while a recent modeling study of the \textit{Ets}-DNA system arrives at
the same conclusion \cite{Wolynes}. The first condition can thus be satisfied as well 
under reasonable conditions. Finally, the measured \textit{lac} repressor binding rates 
\cite{BWH} are comparable to the thermodynamic limiting rate. We conclude that, 
under reasonable conditions, the binding rate of transcription factor proteins can be 
of the same order of magnitude as the thermodynamic limiting rate if the energy spectrum
of conformational fluctuations is determined, under selective pressure, by minimization
of the overall binding time.

We would like to thank Leonid Mirny for useful discussions and the Kavli Institute
of Theoretical Physics and the Aspen Center for Physics, where this study was
initiated, for their hospitality. LH and AG would like to acknowledge support by
the MRSEC Program of the NSF under Award Number DMR-0212302 and RB would like
to acknowledge support by the NSF under DMR Grant 0404507.


\begin{thebibliography}{99}

\bibitem{Alberts} B. Alberts \textit{et al.}, \textit{Molecular Biology of the Cell}, (Garland, New York, 1994).

\bibitem{Adam_Delbruck} G. Adam and M. Delbr\"{u}ck, \textit{Structural Chemistry and Molecular Biology},
A. Rich and N. Davidson, editors, (Freeman, New York, 1968).

\bibitem{Riggs} A. D. Riggs, S. Bourgeois and M. Corn, J. Mol. Biol. \textbf{53}, 401 (1970).

\bibitem{Richter} P. H. Richter and M. Eigen, Biophys. Chem. \textbf{2}, 255 (1974).

\bibitem{BWH} O. G. Berg, R. B. Winter and P. H. von Hippel, Biochemistry. \textbf{20}, 6929 (1981),
R. B. Winter, O. G. Berg and P. H. von Hippel, Biochemistry. \textbf{20}, 6961 (1981),
P. H. von Hippel and O. G. Berg, J. Biol. Chem. \textbf{264}, 675 (1989).

\bibitem{Kalodimos} C. G. Kalodimos \textit{et al.}, Science. \textbf{305}, 386 (2004).

\bibitem{Mirny} M. Slutsky, L. A. Mirny, Biophys. J. \textbf{87}, 4021 (2004).

\bibitem{Wolynes} A specific realization of a conformational fluctuation spectrum of this type
was recently discussed for the \textit{Ets}-DNA system by Y. Levy, J. N. Onuchic and P. G. Wolynes,
J. Am. Chem. Soc. \textbf{129}, 738 (2007).

\bibitem{Radius} More precisely, $b$ is defined as the radius of a cylinder surrounding the DNA duplex
such that a protein will be captured by the DNA, for example by electrostatic attraction, if the 
center of the protein is located inside the cylinder.

\bibitem{Pontryagin} L. Pontryagin, A. Andronov and A. Vitt, Zh. Eksp. Teor. Fiz. 
\textbf{3}, 165 (1933); translated and reprinted in \textit{Noise in Nonlinear Dynamical Systems}. Vol. 1,
edited by F. Moss and P. V. E. McClintock, p. 329, (Cambridge University Press, Cambridge, 1989).

\bibitem{Grosberg} T. Hu, B. I. Shklovskii and A. Y. Grosberg, Biophys. J. \textbf{90}, 2731 (2006).

\bibitem{LambdaP} The maximum size $\lambda_\infty$ of the antenna length is, for the parameter values
used in this paper, less than the DNA persistence length.

\bibitem{Xie} J. Elf, G. Li and X. S. Xie, Science. \textbf{316}, 1191 (2007).

\bibitem{Austin} Y. M. Wang, R. H. Austin and E. C . Cox, Phys. Rev. Lett. \textbf{97}, 048302 (2006).

\bibitem{Huang} Y. Kao-Huang \textit{et al.}, Proc. Natl. Acad. Sci. USA. \textbf{74}, 4228 (1977).



\end{thebibliography}
\end{document}